\documentclass[3p,11pt]{elsarticle}
\usepackage{hyperref}
\journal{Journal}
\usepackage{subcaption}
\usepackage{amsmath}
\usepackage{amssymb}
\usepackage{gensymb}
\usepackage{float}
\usepackage{placeins}
\usepackage{import}
\usepackage{xcolor}
\usepackage{pgf}
\usepackage{soul}
\usepackage{booktabs}
\usepackage[noabbrev,nameinlink]{cleveref}
\crefname{appendix}{}{}

\begin{document}
	

\newcommand{\latticeconstant}{a}                                                
\newcommand{\latticespacing}{d}                                                 
\newcommand{\burgersvector}{b}                                                  
\newcommand{\schmid}{m}                                                         
\newcommand{\rss}{\tau}                                                         
\newcommand{\crss}{s}                                                           
\newcommand{\ratesensitivity}{r}                                                
\newcommand{\hardeningrate}{k_0}                                                
\newcommand{\saturationcrss}{s_\infty}                                          
\newcommand{\hardeningexp}{a}                                                   
\newcommand{\stress}{\sigma}                                                    
\newcommand{\strain}{\varepsilon}												
\newcommand{\youngs}{E}                                                         
\newcommand{\shearmodulus}{G}                                                   
\newcommand{\length}{L}                                                         
\newcommand{\slipangle}{\theta}                                                 
\newcommand{\planeangle}{\phi}                                                  
\newcommand{\displacement}{u}                                                   
\newcommand{\slipstep}{v}                                                       
\newcommand{\slipvelocity}{\dot{\slipstep}}                                     
\newcommand{\refslipvelocity}{\slipvelocity_0}                                  
\newcommand{\sourcelength}{\lambda}                                             
\newcommand{\sourcelengthpdf}{p(\sourcelength)}                                 
\newcommand{\radius}{R}                                                         
\newcommand{\nucleationstress}{s_\text{nuc}}                                    
\newcommand{\minnucleationstress}{\nucleationstress^\text{min}}                 
\newcommand{\maxnucleationstress}{\nucleationstress^\text{max}}                 
\newcommand{\sourcenucleationpdf}{g^+(\nucleationstress)}                         
\newcommand{\nucleationpdf}{p(\nucleationstress)}                               
\newcommand{\nucleationcdf}{P(\nucleationstress)}                               
\newcommand{\minnucleationpdf}{p_\text{min}(\nucleationstress)}                 
\newcommand{\sourcechance}{f_\text{src}}                                        
\newcommand{\edgenucleationpdf}{g^-(\nucleationstress)}                           
\newcommand{\latticefriction}{s_\text{fric}}                                    
\newcommand{\dislocationdensity}{\rho_\text{dis}}                               
\newcommand{\shearstrain}{\gamma}                                               
\newcommand{\defgradient}{\mathbf{F}}                                           
\newcommand{\velgradient}{\mathbf{L}}                                           
\newcommand{\slipdirection}{\vec{s}_0}                                          
\newcommand{\slipplanenormal}{\vec{n}_0}                                        
\newcommand{\mean}{\mu}                                                         
\newcommand{\stddeviation}{\sigma}                                              
\newcommand{\bandwidth}{l}                                                      
\newcommand{\sourcedensity}{\rho_\text{src}}									

\newcommand{\identity}{\mathbf{I}}
\newcommand{\secondpiolaelastic}{\mathbf{S}_e}                                  
\newcommand{\defgrad}{\mathbf{F}}												
\newcommand{\defgradelastic}{\defgrad_e}										
\newcommand{\defgradplastic}{\defgrad_p}										
\newcommand{\defgradplasticdot}{\dot{\defgrad}_p}								
\newcommand{\velgrad}{\mathbf{L}}												
\newcommand{\velgradelastic}{\velgrad_e}										
\newcommand{\velgradplastic}{\velgrad_p}										
\newcommand{\greenlagrange}{\mathbf{E}}                                         
\newcommand{\elasticitytensor}{\mathbb{C}}                                  
\newcommand{\schmidtensor}{\mathbf{P}_0}										
\newcommand{\nonschmidtensor}{\mathbf{P}_\text{NS}} 							
\newcommand{\projectionplanenormal}{\vec{n}{'}_0}								
\newcommand{\nonschmidcoeff}{a}												    
\newcommand{\nonschmidstress}{\tau_{\text{NS}}} 								
\newcommand{\rssyield}{\tau_y}													
\newcommand{\slipfamilyA}{\{110\}\left<111\right>}								
\newcommand{\slipfamilyB}{\{112\}\left<111\right>}								
\newcommand{\slipfamilyC}{\{123\}\left<111\right>}								
\newcommand{\planefamilyA}{\{110\}}
\newcommand{\planefamilyB}{\{112\}}
\newcommand{\planefamilyC}{\{123\}}
\newcommand{\compositefactor}{c}												
\newcommand{\mrsspstress}{\tau^*}												
\newcommand{\crsseff}{\crss^*}
\newcommand{\segmentlength}{L_\text{seg}}										

\begin{frontmatter}

\title{Discrete slip plane analysis of ferrite microtensile tests: On the influence of dislocation source distribution and non-Schmid effects on slip system activity.}

\author{J. Wijnen}

\author{J.P.M. Hoefnagels}

\author{M.G.D. Geers}

\author{R.H.J. Peerlings\corref{peerlings}}
\ead{r.h.j.peerlings@tue.nl}

\address{Department of Mechanical Engineering, Eindhoven University of Technology, 5600~MB~Eindhoven, The~Netherlands}

\cortext[peerlings]{Corresponding author}

\begin{abstract}

The slip system activity in microtensile tests of ferrite single crystals is compared with predictions made by the discrete slip plane model proposed by Wijnen et al. (International Journal of Solids and Structures 228, 111094, 2021). This is an extension of conventional crystal plasticity in which the stochastics and physics of dislocation sources are taken into account in a discrete slip band. This results in discrete slip traces and non-deterministic mechanical behavior, similar to what is observed in experiments. A detailed analysis of which slip systems are presumed to be active in experiments is performed. In small-scale mechanical tests on BCC metals and alloys, non-Schmid effects are often needed to explain the observed response. Therefore, these effects are incorporated into the model by extending a non-Schmid framework commonly used to model $\planefamilyA$ slip to $\planefamilyB$ planes. The slip activity in the simulations is compared to the slip activity in single crystal ferrite microtensile tests. This is done for the discrete slip plane model with and without non-Schmid effects, as well as for a conventional crystal plasticity model. The conventional crystal plasticity model fails to predict the diversity in active slip systems that is observed experimentally. The slip activity obtained with the discrete slip plane model is in convincingly better agreement with the experiments. Including non-Schmid effects only entails minor differences. This suggests that stochastic effects dominate the behavior of ferrite crystals with dimensions in the order of a few micrometers and that non-Schmid effects may not play a large role.
\end{abstract}


\end{frontmatter}



\section{Introduction}

Ferrite ($\alpha$-Fe) is a common phase in many advanced high-strength steels. Improving these steel grades requires accurate modeling of the underlying polycrystalline microstructure, for which a detailed understanding of the plastic behavior of ferrite at the microscale is essential. Small-scale mechanical tests on single crystals, such as micropillar compression tests or microtensile tests, play an important role in unraveling the plastic behavior of metals and alloys and have been extensively used to study both Face-Centered-Cubic (FCC) and Body-Centered-Cubic (BCC) crystals \cite{uchic2009, greer2011, shahbeyk2019}, including ferrite. The BCC structure of ferrite does, however, lead to more intricate behavior compared to FCC phases, such as austenite. Plasticity of BCC metals is governed by the glide of screw dislocations, which have some remarkable properties due to their compact and non-planar core \cite{christian1983, frederiksen2003}. Plastic slip can occur on multiple slip plane families. While at low temperatures slip is mainly observed on $\{110\}$ planes, slip on $\{112\}$ planes is commonly observed at room temperature \cite{franciosi2015, hagen2016, spitzig1970, weinberger2013}. Occasionally, even slip on $\{123\}$ planes is observed \cite{tian2020}. Furthermore, BCC metals display non-Schmid effects, which means that stress components other than the resolved shear stress on a particular slip system affect the activation of dislocation glide \cite{duesberry1998, vitek2004}. Non-Schmid effects give rise to an orientation-dependent mechanical response in small-scale mechanical tests of single ferrite crystals \cite{spitzig1970, rogne2015, hagen2017}. Additionally, they can cause the activation of slip systems that are not among the most favorably oriented systems in the considered test \cite{kheradmand2019}.

Different CP frameworks that are used to study the ferrite response in small-scale mechanical tests incorporate non-Schmid effects \cite{lim2015, mapar2017, joudivand2021}. However, large variations in slip system activity and stress-strain curves are often observed in small-scale mechanical tests of ferrite single crystals \cite{tian2020,du2018}, which can be attributed to the scarcity of dislocation sources in such small volumes \cite{parthasarathy2007, ng2008, elawady2009}. This makes it difficult to determine whether slip on a secondary slip system is due to non-Schmid effects or due to the absence of an easily activated dislocation source on the primary slip system. Conventional CP models do not account for these stochastic effects of the dislocation sources. Furthermore, CP simulations result in strain fields that are much more homogeneous and smooth than the heterogeneous plasticity and localized slip traces observed in experiments.

In this study, the discrete slip plane (DSP) model, introduced by Wijnen et al.\ \cite{wijnen2021}, is adopted to analyze the slip system activity in microtensile tests on ferrite single crystals, performed by Du et al.\ \cite{du2018}. This model accounts for the heterogeneity in plastic deformation due to variations in slip resistances between atomic planes. To determine the slip resistance of atomic planes, dislocation sources are randomly distributed throughout the sample. The strength of these dislocation sources is sampled from a statistical distribution that is based on the physics of dislocation sources in samples with confined dimensions. Since both the spatial and strength distribution of dislocation sources are taken into account, this results in non-deterministic heterogeneous plastic flow within a single crystal. By making some assumptions, the model can be formulated as an enriched continuum crystal plasticity model, which significantly reduces the computational cost and makes it suitable for full-scale simulations of micron-sized mechanical tests.

As part of this work, additional analyses on the slip system activity in the experiments of Du et al.\ \cite{du2018} have been carried out. The qualitative characterization of visible slip traces is extended to quantify the amount of slip on the various active slip systems. This allows for a quantitative comparison between the slip system activity in the experiments and the CP model calibrated as identified \cite{du2018}. Additionally, the results are here compared to the slip system activity in the DSP simulations. Three different parameter sets are used for the DSP simulations. One parameter set only includes Schmid-based glide, while the other two parameter sets include non-Schmid effects through parameters that are identified from experiments \cite{patra2014} or atomistic simulations \cite{chen2013}. Following this approach, the influences of both the stochastic effects and the non-Schmid effects, and their relative importance, are studied.

Non-Schmid effects can be taken into account in crystal plasticity (CP) simulations through extra stress projection tensors. A physically based formulation for $\slipfamilyA$ slip systems, inspired by atomistic simulations, was introduced by Groger and Vitek \cite{groger2008}. This formulation makes use of a secondary projection plane, similar to the works of Qin and Bassani \cite{qin1992a,qin1992b}, and contains three non-Schmid projection terms. A benefit of the three-term formulation is that it is relatively straightforward to characterize from atomistic simulations \cite{groger2008, koester2012, chen2013}. Additionally, it was recently shown that the three-term formulation maybe recovered from ab-initio calculations, which allows for a physical interpretation of the parameters \cite{kraych2019, clouet2021}. The 5-term non-Schmid formulation, introduced by Asaro and coworkers \cite{asaro1977, dao1993, dao1996}, is also adopted in several studies \cite{yalcinkaya2008, lim2013, knezevic2014}. This formulation is often referred to as purely phenomenological and more general. However, Gr\"oger and Vitek \cite{groger2019} recently showed that the extra terms in this formulation arise because of the unfortunate choice of the auxiliary projection plane and that both formulations are actually equivalent. Therefore, the three-term formulation will be incorporated into the DSP model. However, the original three-term formulation is only defined for $\slipfamilyA$ slip systems, while slip on $\slipfamilyB$ slip systems is frequently observed in the analyzed experiments. Therefore, the three-term formulation is extended to $\slipfamilyB$ slip systems. This is done based on the assumption of composite slip \cite{marichal2013}.

The structure of this paper is as follows. In \Cref{section:experiments} the experiments done by Du et al.\ \cite{du2018} are briefly summarized, as is the procedure for accurately analyzing the slip system activity. The numerical model is presented in \Cref{section:model}. This includes a review of the most important aspects of the DSP model presented in Wijnen et al.\ \cite{wijnen2021} and the extension of the three-term non-Schmid formulation to $\slipfamilyB$ slip systems. Additionally, the distribution used for the stochastics of dislocation sources and the adopted simulation parameters are discussed. In \Cref{section:results} the results of the experimental and numerical analysis are presented and discussed. Finally, the conclusions are summarized in \Cref{section:conclusions}.


\subsection*{Nomenclature}

\FloatBarrier
\begin{table*}[h]
	\begin{tabular}{l@{\hskip 4em}l}
		$a$											  & scalar \\
		$\vec{a}$                                     & vector \\
		$\mathbf{A}$                                  & second-order tensor         \\
		$\mathbb{A}$                                  & fourth-order tensor         \\
		$\mathbf{C} = \vec{a} \otimes \vec{b}$        & dyadic product              \\
		$\mathbf{C}=\mathbf{A} \cdot \mathbf{B}$      & single tensor contraction \\
		$c = \mathbf{A} : \mathbf{B} = A_{ij} B_{ji}$ & double tensor contraction \\
		$\vec{c} = \vec{a} \times \vec{b}$ 	  		  & cross product
		
	\end{tabular}
\end{table*}
\FloatBarrier

\section{Analysis of the experimental data}
\label{section:experiments}

\subsection{Sample characterization}

The uniaxial microtensile tests on interstitial free ferrite performed by Du et al.\ \cite{du2018} are analyzed in more detail here, to facilitate an in-depth comparison with predictions made by the modeling. The length of the gauge section of these specimens is 9 $\mu$m. The cross section is rectangular with dimensions $3\times2$ $\mu$m. The dislocation density of the specimen has been measured to be approximately $7\cdot10^{11}$ m$^{-2}$. 

A total of thirteen specimens were cut from three different grains with different crystal orientations. The Euler angles of the three grains are given in \Cref{tab:eulerangles} in \Cref{appendix:slipsystems}. In \Cref{fig:slipdirections} the slip directions together with the $\planefamilyA$ and $\planefamilyB$ plane normals of the grains are plotted in a pole figure, where the projection plane normal is equal to the loading direction ($x$-axis). The four slip directions (marked by circles) are denoted by the characters \emph{A}, \emph{B}, \emph{C}, and \emph{D}. A slip direction that makes an angle of $45\degree$ with the loading axis would be ideal for slip, since a slip system with this slip direction may have a Schmid factor of 0.5. Therefore, the slip directions are ordered based on how close their orientation is to $45\degree$ (marked with a green circle in the polar plots). In all three grains, the primary slip system has a slip direction \textit{A}. For Grain 1, all five microtensile tests were analyzed, while all four tests are analyzed for both Grains 2 and 3. A more detailed description of the experiments, e.g.\ regarding sample preparation, nanoforce tensile testing, and scanning electron microscopy (SEM) imaging, can be found in Du et al.\ \cite{du2017,du2018}.

\begin{figure}[tb!]
	\centering
	\begin{subfigure}{\linewidth}
		\centering
		\includegraphics[width=.7\linewidth]{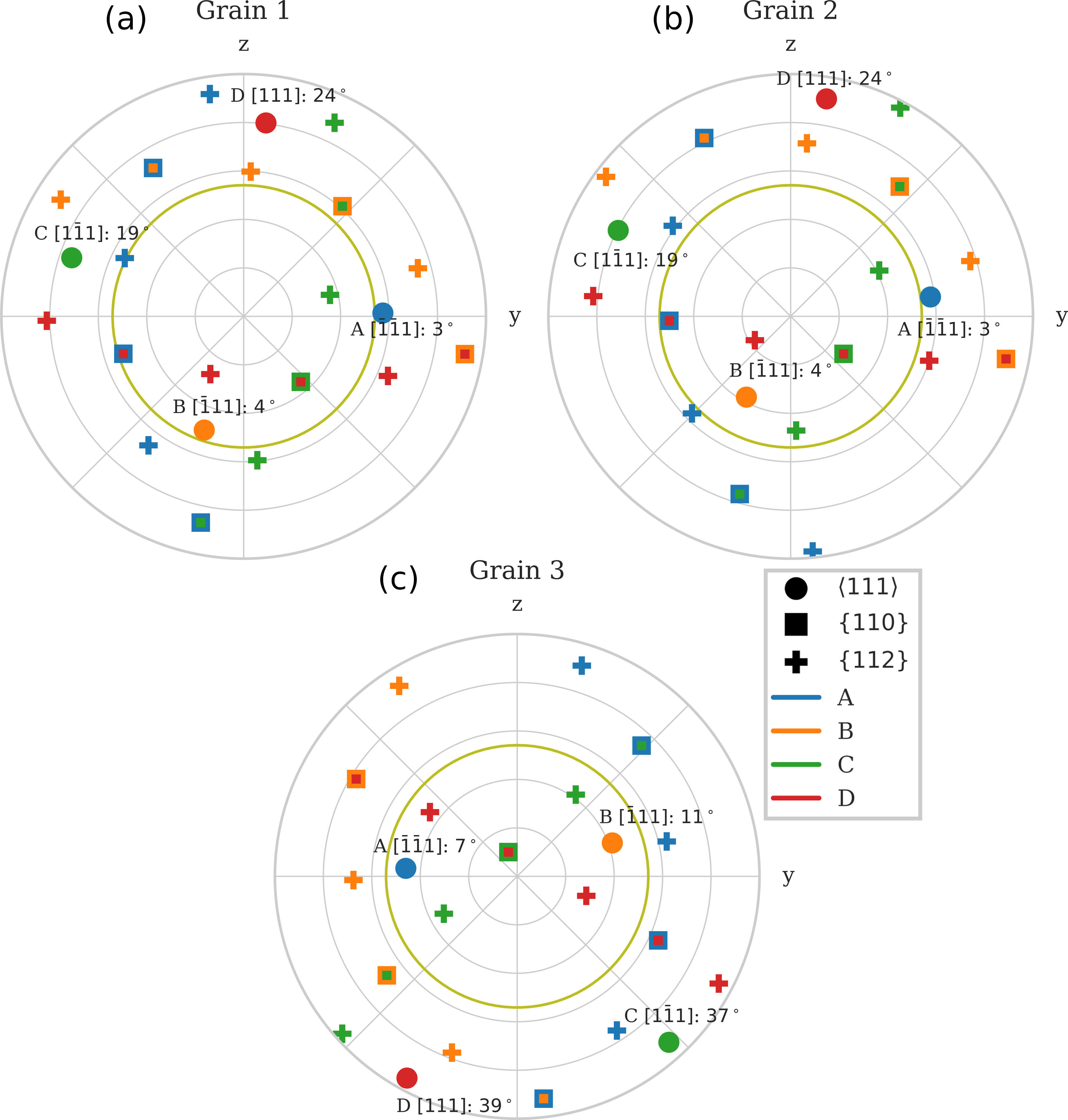}
	\end{subfigure}
	\caption{Polar plots of the $\left< 111 \right>$ slip directions, $\planefamilyA$ plane normals and $\planefamilyB$ plane normals of all three grains. The projection plane normal is parallel to the loading direction ($x$-axis). The slip directions (circles) are denoted by characters \emph{A} through \emph{D}, based on their deviation from the circle that makes a 45$\degree$ angle with the loading axis, marked with a green line in the polar plots. Additionally, the deviation from the 45$\degree$ circle for the slip directions is given in degrees. All slip plane normals belonging to the same slip direction are marked with the same color.}
	\label{fig:slipdirections}
\end{figure}

\subsection{Analysis of slip system activity}
\label{section:exp_analysis}

\begin{figure}[tb!]
	\centering
	\begin{subfigure}{\textwidth}
		\centering
		\includegraphics[width=\linewidth]{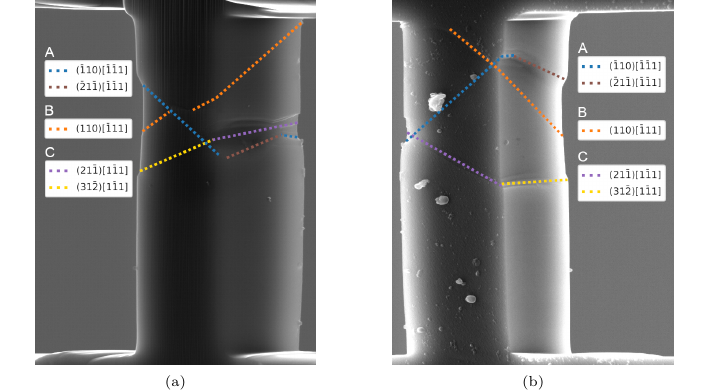}
		\phantomcaption \label{fig:sliptraces_a}
		\phantomcaption \label{fig:sliptraces_b}
	\end{subfigure}
	\caption{Two SEM images of a sample from Grain 1. (a) shows the front and right side of the sample while (b) shows the back and left side of the sample. The identified traces are marked with colors, with a separate color for each slip system.}
	\label{fig:sliptraces}
\end{figure}

A detailed slip system analysis was performed on the samples. Secondary electron images of four different viewpoints were used to identify slip traces as well as the underlying slip systems. These images show the front/right, front/left, back/right, and back/left surfaces of the samples. An example is shown in \Cref{fig:sliptraces}, where the front/right (\Cref{fig:sliptraces_a}) and back/left (\Cref{fig:sliptraces_b}) images of a sample taken from Grain 1 are shown. Images of other samples show similar slip patterns.

The orientations of the slip traces were measured and compared to the theoretical traces of the available slip systems. Note that most slip traces are visible on all faces of the sample. However, a single trace cannot always be assigned to a single slip system. Even on one side of the sample, a trace sometimes has two parts that differ in orientation, for example, the $(\bar110)[\bar1\bar11]$-$(\bar21\bar1)[\bar1\bar11]$ slip trace in \Cref{fig:sliptraces} (respectively the blue and brown dotted lines). This is an indication of cross slip. Nevertheless, a complete slip trace can always be assigned to a specific slip direction, since screw dislocations only change their slip plane when cross slipping while their slip direction is maintained. This means that all parts of a complete slip trace are assigned to slip systems with a common slip direction. In this way, the slip direction of a trace was determined with high certainty. However, occasionally the orientation of part of a slip trace was still close to the theoretical slip trace of two candidate slip systems. 

Consider again the $(\bar110)[\bar1\bar11]$-$(\bar21\bar1)[\bar1\bar11]$ slip trace in \Cref{fig:sliptraces} (respectively the blue and brown dotted lines). Different segments of this trace were assigned to either the $(\bar110)$ plane or the $(\bar21\bar1)$ plane, but all parts have a common $[\bar1\bar11]$ slip direction, i.e.\ slip direction \emph{A}. Furthermore, a slip trace of which segments were assigned to the $(31\bar2)$ plane is visible (yellow dotted line). Occasionally, a $\slipfamilyC$ slip system was required to identify a complete trace with the same slip direction. However, $\planefamilyC$ slip appeared to be significantly less frequent than $\planefamilyA$ and $\planefamilyB$ slip. For this reason, and due to the lack of knowledge on non-Schmid effects for $\planefamilyC$ slip, only the $\slipfamilyA$ and $\slipfamilyB$ slip systems were taken into account in the simulations.

The identified slip traces in the SEM images and the fractional assignment of the slip steps to the active slip systems are presented in the supplementary material for all 13 specimens (5 of grain 1, 4 of grain 2, 4 of grain 3).

\begin{figure}[tb!]
	\centering
	\includegraphics[width=.4\linewidth]{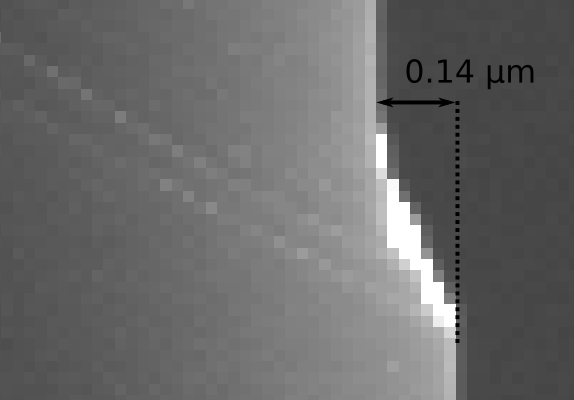}
	\caption{SEM image of a slip step on the back side of a sample. The horizontal length of the slip step in the image is determined to be 0.14 $\mu$m.}
	\label{fig:slipmeasure}
\end{figure}

The amount of slip that has occurred on a particular slip system was determined by using images of the front and/or back of the sample. In \Cref{fig:slipmeasure} the slip step created by the trace of slip direction \emph{C} in \Cref{fig:sliptraces}, as observed on the back side of the sample, is shown. The horizontal length of this slip step was measured to be 0.14 $\mu$m. By projecting the horizontal slip step onto the slip direction of this trace, the total slip step was calculated to be 0.17 $\mu$m. This was done for all clearly visible slip traces in a sample. The normalized slip magnitude in a certain slip direction was then calculated by adding up the measured slip steps of traces with a common slip direction and dividing it by the total slip of all traces measured in this way. For example, the normalized slip magnitudes of the sample shown in \Cref{fig:sliptraces} are $\emph{A}=0.51$, $\emph{B}=0.18$, and $\emph{C}=0.31$. These normalized slip magnitudes can be calculated either per sample or per grain.

To quantitatively determine the slip activity on specific slip systems, the measured slip steps for different slip directions were subdivided. This was done based on the length of the visible traces. The slip trace of slip direction \emph{C} in \Cref{fig:sliptraces} is considered as an example. Two full sides of this slip trace were assigned to the $(21\bar1)[1\bar11]$ and the $(31\bar2)[1\bar11]$ slip systems. Therefore, the amount of slip assigned to these slip systems was half of the total measured slip step of slip trace \emph{C}, i.e.\ 0.085 $\mu$m. 

The analysis and data of all samples is provided in the supplementary material.

\section{Discrete slip plane model}
\label{section:model}

\subsection{Model equations}

The most relevant equations of the discrete slip plane model are explained below. For a more detailed treatment of the model, the reader is referred to \cite{wijnen2021}.

In its basic form, the model considers all atomic slip planes of a specific slip system. The initial slip resistance, $\crss_0$, of a plane is given by
\begin{equation}
	\crss_0 = \nucleationstress + \latticefriction + 0.5 \shearmodulus \burgersvector \sqrt{\dislocationdensity},
	\label{eq:initialcrss}
\end{equation}
where $\nucleationstress$ is the nucleation stress, $\latticefriction$ is the lattice friction, $\shearmodulus$ is the shear modulus, $\burgersvector$ is the length of the Burgers vector and $\dislocationdensity$ is the initial dislocation density. The nucleation stress is assumed to vary significantly per atomic slip plane due to the presence, or absence, of dislocation sources or obstacles. Therefore, its value for each plane is sampled from a probability density function $\nucleationpdf$. This results in a heterogeneous flow stress in a single crystal. The probability density function for the nucleation stress is based on the physics of single-arm dislocation sources, as the plasticity in small specimens is mainly governed by single-arm dislocation sources, as motivated and elaborated in \Cref{section:stochastics}.

Resolving each individual atomic slip plane in simulations is computationally prohibitive. Therefore, in the finite element implementation, the discrete planes are grouped into parallel bands of thickness $\bandwidth$, which typically is much larger than the atomic spacing. This is done for all slip systems. The slip resistance of a band is then taken equal to the slip resistance of the weakest atomic plane contained in that band, i.e.\ the atomic plane with the lowest slip resistance. This essentially results in a crystal plasticity model in which some of the heterogeneity of the (dislocated) crystal, and its response, is preserved. The numerical solution procedure is consistent with a standard crystal plasticity finite element (CPFE) framework and requires finite elements which are sufficiently small to resolve the slip system bands - typically $\bandwidth/2$.

The plastic deformation is accounted for through the multiplicative split of the deformation gradient tensor into an elastic part, $\defgradelastic$, and a plastic part, $\defgradplastic$:
\begin{equation}
	\defgrad = \defgradelastic \cdot \defgradplastic.
\end{equation}

The elastic response of the material is modeled with a Saint Venant-Kirchhoff type of model:
\begin{equation}
	\secondpiolaelastic = \tfrac{1}{2}\elasticitytensor:( \defgradelastic^T \cdot \defgradelastic - \identity ),
\end{equation}
where $\elasticitytensor$ is the fourth-order elasticity tensor.

The kinetics of a certain slip system $\alpha$ is described by a rate-dependent viscoplastic formulation:
\begin{equation}
	\dot{\shearstrain}^\alpha = \frac{\refslipvelocity}{\bandwidth} \left( \frac{|\rss^\alpha|}{\crss^\alpha-\nonschmidstress^\alpha} \right)^\frac{1}{\ratesensitivity} \text{sign}(\rss^\alpha),
	\label{eq:kinetic}
\end{equation}
where $\dot{\shearstrain}^\alpha$ is the average shear strain rate in the band resulting from the slip system activity, $\refslipvelocity$ is a reference slip velocity, $\ratesensitivity$ is a rate sensitivity parameter and $\bandwidth$ is the thickness of a band. The resolved shear stress, $\rss^\alpha$, in the stress-free intermediate configuration can be calculated by
\begin{equation}
	\rss^\alpha = \left( \secondpiolaelastic \cdot \defgradelastic^T \cdot \defgradelastic \right) : \schmidtensor^\alpha,
\end{equation}
where $\secondpiolaelastic$ is the second Piola-Kirchhoff stress tensor defined in the stress-free intermediate configuration. The Schmid tensor, $\schmidtensor^\alpha$, is given by
\begin{equation}
	\schmidtensor^\alpha = \slipdirection^\alpha \otimes \slipplanenormal^\alpha,
	\label{eq:schmidtensor}
\end{equation}
where $\slipdirection$ and $\slipplanenormal$ are the slip direction and slip plane normal in the undeformed (or intermediate) configuration, respectively. Additionally, a non-Schmid stress $\nonschmidstress$ is introduced in \Cref{eq:kinetic} which is not present in the original formulation of the model. This non-Schmid stress is elaborated in \Cref{section:nonschmid}.

The crystallographic decomposition defines the plastic velocity gradient tensor as the following summation over all slip systems:
\begin{equation}
	\velgradplastic = \defgradplasticdot \cdot \defgradplastic^{-1} = \sum_{\alpha=1}^\text{N} \dot{\shearstrain}^\alpha \slipdirection^\alpha \otimes \slipplanenormal^\alpha,
\end{equation}
where $N$ is the number of slip systems taken into account. Finally, the evolution of the slip resistance of a slip system is given by
\begin{equation}
	\dot{\crss}^\alpha = \hardeningrate \bandwidth \left( 1 - \frac{\crss^\alpha}{\saturationcrss} \right)^\hardeningexp \sum_{\beta=1}^N q^{\alpha\beta} |\dot{\shearstrain}^\beta|,
	\label{eq:hardening}
\end{equation}
where $\hardeningrate$ is the initial hardening rate, $\saturationcrss$ is the saturation slip resistance, $\hardeningexp$ is the hardening exponent and $q$ is the cross-hardening matrix, where $\alpha$ and $\beta$ are used to denote a particular slip system.

\subsection{Modeling Non-Schmid effects on $\slipfamilyA$ slip systems}
\label{section:nonschmid}

Non-Schmid effects can be taken into account in a continuum setting by introducing extra stress projection tensors \cite{dao1993,qin1992a,bassani2001}. By collecting multiple non-Schmid projection tensors in a single tensor $\nonschmidtensor^\alpha$, the contribution of non-Schmid effects to plastic slip on a slip system can be written as
\begin{equation}
	\nonschmidstress^\alpha = \left( \secondpiolaelastic \cdot \defgradelastic^T \cdot \defgradelastic \right) : \nonschmidtensor^\alpha
\end{equation}
The non-Schmid stress, $\nonschmidstress^\alpha$, affects the resistance to plastic slip, i.e.\ the glide stress that is required to initiate the motion of dislocations. It is hence introduced in the denominator of \Cref{eq:kinetic} such that it reduces the effective slip resistance, following the approach of Mapar et al.\ \cite{mapar2017}.

The formulation with three non-Schmid projection terms by G\"oger et al.\ \cite{groger2008} is adopted here. The non-Schmid projection tensor of a $\{110\}[111]$ slip system in this formulation is given by
\begin{equation}
	\nonschmidtensor^\alpha = \nonschmidcoeff_1 \slipdirection^\alpha \otimes \projectionplanenormal^\alpha
	+ \nonschmidcoeff_2 \left( \slipplanenormal^\alpha \times \slipdirection^\alpha \right) \otimes \slipplanenormal^\alpha
	+ \nonschmidcoeff_3 \left( \projectionplanenormal^\alpha \times \slipdirection^\alpha \right) \otimes \projectionplanenormal^\alpha,
	\label{eq:nonschmidtensor}
\end{equation}
where $\nonschmidcoeff_1$, $\nonschmidcoeff_2$ and $\nonschmidcoeff_3$ are material parameters, $\slipdirection^\alpha$ is the slip direction, $\slipplanenormal^\alpha$ is the slip plane normal and $\projectionplanenormal^\alpha$ is the normal of a secondary projection plane which is obtained by rotating the slip plane normal over an angle of $-60\degree$ around the slip direction.

The first term in \Cref{eq:nonschmidtensor}, multiplied by material parameter $\nonschmidcoeff_1$, is modeling the twinning/anti-twinning (T/AT) behavior. Initially, T/AT effects were believed to arise due to the asymmetry of the generalized stacking fault energy surface of $\planefamilyB$ planes along the $\left< 111 \right>$ directions \cite{duesberry1998, frederiksen2003}. This makes shearing along the positive (twinning) $[111]$ direction of a $\planefamilyB$ plane easier than shearing along the negative (anti-twinning) $[111]$ direction. However, more recent ab-initio calculations link the T/AT behavior to the deviation of the screw dislocation trajectory from a straight path between equilibrium positions on $\planefamilyA$ planes \cite{dezerald2016}. In this case, parameter $\nonschmidcoeff_1$ gets a physical interpretation and can be connected to the angle of deviation from the straight path.

The other two non-Schmid coefficients, $\nonschmidcoeff_2$ and $\nonschmidcoeff_3$, are modeling the effect of non-glide stresses on the screw dislocation core. These stresses do not contribute to the Peach-Koehler force on the dislocation but do affect the stress required to activate dislocation glide. An extension of the ab-initio modeling methodology used to investigate the T/AT effects shows that the effect of non-glide stresses can be explained by the anisotropic variation of the dilatation of a gliding screw dislocation and its elastic coupling to the applied stress \cite{kraych2019}. The stresses related to parameters $\nonschmidcoeff_2$ and $\nonschmidcoeff_3$ are resolved shear stresses on a plane that contains the slip direction but in a direction perpendicular to the slip direction. The vectors $\slipdirection$, $\slipplanenormal$ and $\projectionplanenormal$ of the $\{110\}[111]$ slip systems are given in \Cref{tab:slipfamilyA} in \Cref{appendix:slipsystems}. Note that a different secondary projection plane should be used for positive slip than for negative slip on a slip system, i.e.\ $\projectionplanenormal$ depends on the sign of $\rss^\alpha$. Additionally, note that the use of \Cref{eq:nonschmidtensor} is limited to $\slipfamilyA$ slip systems since the secondary projection planes with normals $\projectionplanenormal$ are only defined for $\planefamilyA$ planes.

\subsection{Modeling Non-Schmid effects on $\slipfamilyB$ slip systems}

The framework described in the previous section has been used to model non-Schmid effects on $\slipfamilyA$ slip systems in various BCC metals. Slip systems of the $\slipfamilyB$ family are usually not considered. One of the primary reasons for this is that molecular dynamics simulations (with interatomic potentials that result in non-degenerate or compact screw dislocation cores, which is assumed to be the correct core structure based on more accurate ab-initio calculations \cite{frederiksen2003, ventelon2010, clouet2021}) only display slip on $\{110\}$ planes. However, slip on $\{112\}$ planes is frequently observed in experiments, especially for ferrite.

\begin{figure}[!tb]
	\centering
	\includegraphics[width=.3\linewidth]{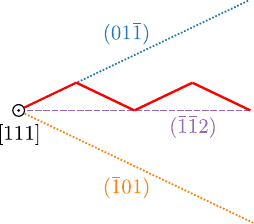}
	\caption{Sketch of composite zig-zag slip on a $\planefamilyB$ slip plane. Alternating slip steps of a screw dislocation on the $(01\bar1)$ and $(\bar101)$ planes result in an apparent slip on the $(\bar1\bar12)$ plane. The direction of the Burgers vector of the screw dislocation is $\left[111\right]$, which is pointing out of the plane of the sketch.}
	\label{fig:compositeslip}
\end{figure}

The precise mechanism behind slip on the $\slipfamilyB$ family is poorly understood. Interestingly, in early molecular dynamics (MD) simulations with potentials resulting in degenerate screw dislocation cores, slip was observed on $\{112\}$ planes only \cite{chaussidon2006, wen2000, marian2004}. This degenerate core has two variants due to asymmetry, entailing zig-zag slip. One variant only propagates on a specific $\{110\}$ plane by kink pair nucleation. After an elementary step, the core structure switches to the other variant. This variant only propagates on another $\{110\}$ plane with the same Burgers vector. The resulting composed slip plane of these alternating elementary steps is a $\{112\}$ plane with that same Burgers vector. This is schematically shown in \Cref{fig:compositeslip}, where a screw dislocation with a $\left<111\right>$ Burgers vector, pointing out of the plane of the sketch, is depicted. By alternating slip steps of the same size on the $(01\bar1)$ and $(\bar101)$ planes, the apparent slip plane becomes $(\bar1\bar12)$.

The hypothesis that slip on $\slipfamilyB$ slip systems consists of zig-zag slip on two $\slipfamilyA$ slip systems is also made in earlier literature \cite{noble1965,sestak1965,christian1983}. More recently, Marichal et al. \cite{marichal2013} investigated slip planes in Tungsten single crystal compression tests with in-situ Laue diffraction and confirmed that slip on apparent $\planefamilyB$ planes can be described as composed zig-zag slip of elementary steps on $\planefamilyA$ planes, as long as both $\planefamilyA$ planes are equally stressed.

The hypothesis of composite slip is adopted here. In that case, every slip plane of the $\slipfamilyB$ family can be written as a combination of two slip planes of the $\slipfamilyA$ family. For example:
\begin{equation}
	(\bar1\bar12)=(\bar101)-(01\bar1).
\end{equation}
Instead of calculating the Schmid tensor of, for instance, the $(\bar1\bar12)\left[111\right]$ system through \Cref{eq:schmidtensor}, it is now written as a combination of the Schmid tensors of the $(01\bar1)\left[111\right]$ and $(\bar101)\left[111\right]$ slip systems:
\begin{equation}
	\schmidtensor^{(\bar1\bar12)} = \compositefactor \left( \schmidtensor^{(\bar101)} - \schmidtensor^{(01\bar1)} \right),
	\label{eq:schmidcomposite}
\end{equation}
where $\compositefactor$ is the composite slip factor. When a value of $1/\sqrt{3}$ is adopted for $\compositefactor$, the obtained Schmid tensor is equal to the one obtained by \Cref{eq:schmidtensor}, which means that slip due to pure glide stresses will occur just as easily on $\planefamilyB$ planes as on $\planefamilyA$ planes if both slip plane families have the same slip resistance. Slip on $\planefamilyB$ planes becomes more difficult for a lower value of $\compositefactor$ and less difficult for a higher value of $\compositefactor$. Here, the value of $1/\sqrt{3}$ is adopted since this corresponds with the conventional way of modeling slip on $\slipfamilyB$ slip systems through \Cref{eq:schmidtensor}. The combinations of $\slipfamilyA$ slip systems which result in $\slipfamilyB$ slip systems are given in \Cref{tab:slipfamilyB} in \Cref{appendix:slipsystems}.

Adopting a similar approach allows for the calculation of the non-Schmid projection tensors of the $\slipfamilyB$ slip systems. For the $(\bar1\bar12)\left<111\right>$ slip system the non-Schmid tensor becomes
\begin{equation}
	\nonschmidtensor^{(\bar1\bar12)} = \compositefactor \left( \nonschmidtensor^{(\bar101)} - \nonschmidtensor^{(01\bar1)} \right).
	\label{eq:nonschmidcomposite}
\end{equation}

\subsection{The effect of non-Schmid stresses on dislocation glide}

\begin{figure}[tb!]
	\begin{subfigure}{\linewidth}
		\centering
		\includegraphics[width=\linewidth]{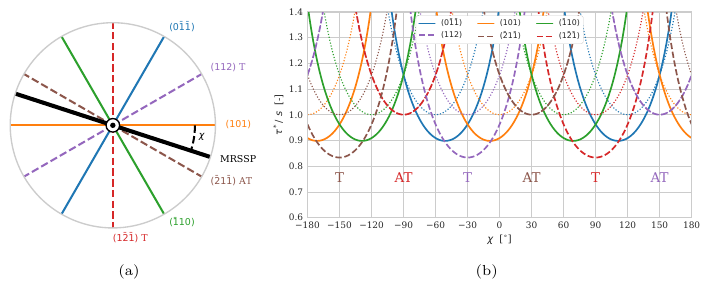}
		\phantomcaption \label{fig:mrssp}
		\phantomcaption \label{fig:mrsspactivation}
	\end{subfigure}
	\caption{(a) Schematic of a $\left[111\right]$ screw dislocation with $\planefamilyA$ and $\planefamilyB$ slip planes. A shear stress is applied over the MRSSP, which is defined by its angle, $\chi$, with respect to the $(101)$ plane. (b) The ratio between the shear stress applied to the MRSSP, $\mrsspstress$, and the slip resistance, $\crss$, for the case where $\mrsspstress/(\crss-\nonschmidstress)=1$, as a function of the MRSSP angle $\chi$. The dislocation considered is a $[111]$ screw dislocation. Twinning effects are taken into account by setting $\nonschmidcoeff_1=0.2$. The dotted lines represent the case where only Schmid glide is considered ($\nonschmidcoeff_1=0$).}
\end{figure}

To demonstrate the effect of non-Schmid stresses on screw dislocation mobility, the straight $[111]$ screw dislocation depicted in \Cref{fig:mrssp} is considered. The Burgers vector points out of the plane of the sketch. A shear stress $\mrsspstress$ is applied on a plane containing the dislocation in a direction parallel to the dislocation line. This plane is termed the maximum resolved shear stress plane (MRSSP) in the remainder of this section. The direction of this shear stress is parallel to the Burgers vector, i.e.\ also pointing out of the figure. The MRSSP is rotated with an angle $\chi$ with respect to the $(101)$ plane. The ratio $\mrsspstress/\crss$ at yield is plotted for all $\slipfamilyA$ and $\slipfamilyB$ slip systems as a function of $\chi$ in \Cref{fig:mrsspactivation}, i.e.\ the lines indicate the values where $\mrsspstress/(\crss-\nonschmidstress)=1$, which marks the activation of slip systems. The lower envelope of these curves may be regarded as a yield surface for this particular applied shear stress and the lowest curve for a given angle $\chi$ indicates which system is activated first at that angle - and at which level of stress. The dotted lines represent the case where only Schmid stresses are considered ($\nonschmidcoeff_1=\nonschmidcoeff_2=\nonschmidcoeff_3=0$). Here, the minimum value of $\mrsspstress/\crss$ is equal to 1 for each slip system. Moreover, the location of this minimum always corresponds to the angle, $\chi$, for which the MRSPP coincides with the slip plane of that particular slip system. 

The solid lines represent the $\slipfamilyA$ slip systems where T/AT effects are taken into account but omitting the effects of non-glide stresses ($\nonschmidcoeff_1=0.2$, $\nonschmidcoeff_2=\nonschmidcoeff_3=0$). Here, the minimum value of $\mrsspstress/\crss$ is no longer located at the value of $\chi$ at which the MRSSP is coinciding with the slip plane. Instead, the minimum shifts towards a $\planefamilyB$ twinning plane. 

The dashed lines represent the $\slipfamilyB$ slip systems with only T/AT effects taken into account. Here, the location of the minimum of $\mrsspstress/\crss$ is still located at the value of $\chi$ where the MRSSP is coinciding with the slip plane. However, there is a significant difference between the minimum value of $\planefamilyB$ twinning directions and anti-twinning directions, i.e.\ slip occurs much easier in twinning directions. Furthermore, a rotation of $180\degree$ of the MRSSP, equivalent to changing the sign of $\mrsspstress$, changes $\planefamilyB$ twinning directions to anti-twinning directions and vice versa. This T/AT behavior of $\slipfamilyB$ slip systems corresponds to what usually is observed in experiments, which justifies the extension of the three-term non-Schmid formulation to $\slipfamilyB$ slip systems through \Cref{eq:nonschmidcomposite} for T/AT effects. 

The effect of the non-glide stresses cannot be demonstrated by only applying the shear stress, $\mrsspstress$, to the MRSSP. Instead, an additional shear stress perpendicular to $\mrsspstress$ should be applied to the MRSSP. When a constant value is taken for this perpendicular shear stress and one of the non-glide stresses is taken into account, the result is similar to that of \Cref{fig:mrsspactivation}: slip on half of the $\planefamilyB$ planes becomes easier and the minima of the $\planefamilyA$ planes shift towards the easier $\planefamilyB$ planes.

\subsection{Nucleation stress stochastics}
\label{section:stochastics}

The stress required to nucleate new dislocations, $\nucleationstress$, varies significantly between
atomic planes. On planes containing a dislocation source, the nucleation stress is relatively low. Contrarily, on planes that do not contain a source, dislocations have to be nucleated from the free surface, which results in a much higher nucleation stress. Therefore, the probability density function from which the nucleation stresses of the planes are sampled consists of two contributions, as explained in detail in \cite{wijnen2021}:
\begin{equation}
	\nucleationpdf = \sourcechance \sourcenucleationpdf + (1-\sourcechance) \edgenucleationpdf,
	\label{eq:nucleationpdf}
\end{equation}
where $\sourcenucleationpdf$ is the probability density function associated with planes that contain a dislocation source,  and $\edgenucleationpdf$ for those without a source, $\sourcechance$ is the probability that a plane contains a dislocation source, which can be estimated by
\begin{equation}
	\sourcechance = \sourcedensity A \latticespacing,
	\label{eq:sourcechance}
\end{equation}
with $A$ the area of a lattice plane in the sample, $\latticespacing$ the lattice spacing and $\sourcedensity$ the dislocation source density. 

Because the mechanical behavior of the samples considered in our study is dominated by a few dislocation sources, the choice of $\edgenucleationpdf$ generally has a negligible effect. Therefore, a narrow normal distribution around the theoretical strength is adopted (see \cite{wijnen2021}). On the contrary, the choice for $\sourcenucleationpdf$ in \Cref{eq:nucleationpdf} is important. It has been shown that the single-arm (SA) dislocation source model introduced by Parthasarathy et al.\ \cite{parthasarathy2007} is capable of predicting the size-dependent strength of micropillars of both FCC \cite{ng2008,pan2015,takeuchi2021,wijnen2021} and BCC \cite{lee2012,soler2014,yilmaz2019} metals and alloys. The model assumes that dislocation sources in such small samples are dislocations that are pinned at one end, while the other end is connected to the free surface. When a stress is applied, this dislocation spins around its pinning point, resulting in plastic slip on its slip plane. The stress required to operate a SA source is inversely related to the shortest distance to the free surface, $\sourcelength$, since this configuration results in the largest curvature and, consequently, line tension of the dislocation \cite{rao2007}.

In the single-arm source model, every dislocation segment can act as a possible dislocation source. Therefore, following Parthasarathy et al.\ \cite{parthasarathy2007} and later studies \cite{ng2008,soler2014}, the source density, $\sourcedensity$, in \Cref{eq:sourcechance} can be estimated from the initial dislocation density, $\dislocationdensity$, by
\begin{equation}
	\sourcedensity = \frac{\dislocationdensity}{N\segmentlength},
\end{equation}
where $N$ is the number of slip systems and $\segmentlength$ is the average length of dislocation segments in the specimen, taken as the sample radius.

\begin{figure}[tb!]
	\centering
	\begin{subfigure}[b]{.9\linewidth}
		\centering
		\includegraphics[width=\linewidth]{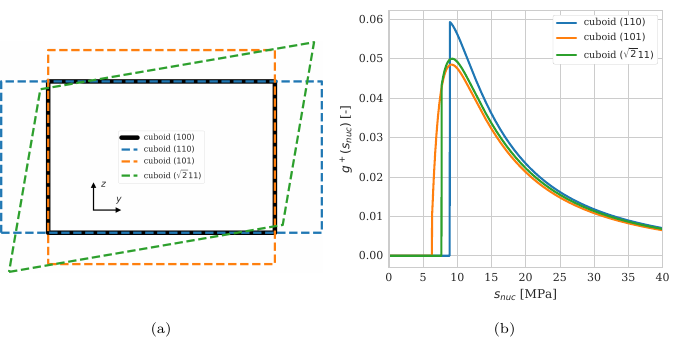}
		\phantomcaption \label{fig:shapes}
		\phantomcaption \label{fig:distributions}
	\end{subfigure}
	\caption{(a) Parallelogram-shaped slip planes in a cuboid sample. The black solid shape has a normal vector equal to $(100)$, i.e.\ it is aligned with the loading direction. The normal vectors of the other three planes shown are all at an angle of 45$\degree$ with the loading axis. (b) Probability density functions of the nucleation stress for the three differently oriented slip planes in (a).}
\label{fig:planes}
\end{figure}

The probability density function of the shortest distance to the free surface, $\sourcelength$ for a pinning point with uniform position distribution, has previously been formulated for ellipsoidal-shaped planes in cylindrical specimens \cite{parthasarathy2007}. However, this study considers samples with a cuboid-shaped gauge section. A slip plane in these cuboid specimens has the shape of a parallelogram. Examples of such parallelograms in a cuboid geometry with dimensions $L_y=3$ $\mu$m and $L_z=2$ $\mu$m are depicted in \Cref{fig:shapes}. Four different slip planes with different normal vectors are shown. Using geometrical considerations, the probability density function of $\sourcelength$ for uniformly distributed pinning points for these planes is derived as
\begin{equation}
	g^+(\sourcelength) = \frac{2 \, C_A}{C_B} - \frac{8 \sourcelength}{C_B \, C_C},
	\label{eq:sa_cuboid}
\end{equation}
with
\begin{equation}
	C_A = L_y \sqrt{\left(\frac{n_y}{n_x}\right)^2 + 1} + L_z \sqrt{\left(\frac{n_z}{n_x}\right)^2 + 1},
\end{equation}
\begin{equation}
	C_B = L_y L_z \sqrt{\left(\frac{n_y}{n_x}\right)^2 + \left(\frac{n_z}{n_x}\right)^2  + 1}
\end{equation}
and
\begin{equation}
	C_C = \sqrt{1 - \frac{n_y^2n_z^2}{\left(n_x^2 + n_y^2\right)\left(n_x^2 + n_z^2\right)}},
\end{equation}
where $L_y$ and $L_z$ are the edge lengths of the specimen in the $y-$ and $z-$direction, i.e.\ $L_y=3$ $\mu$m and $L_z=2$ $\mu$m in the tested samples, and $n_x$, $n_y$ and $n_z$ are the components of the slip plane normal $\slipplanenormal$ with unit length. Note that the above formulation assumes that the $x$-axis is aligned with the central axis of the specimen.

In \Cref{fig:distributions}, the distributions for $\nucleationstress$ based on \Cref{eq:sa_cuboid} are plotted for slip planes with normal vectors $(110)$, $(101)$ and $(\sqrt211)$. All three normal vectors represent an area at an angle of 45 $\degree$ with respect to the central $x$-axis (i.e.\ the $[100]$-loading direction). However, the angle over which they are rotated around the $x$-axis is different. This results in differently shaped slip planes (\Cref{fig:shapes}), and hence in different distributions, as can be seen in \Cref{fig:distributions}.


\section{Simulation results}
\label{section:results}

In this section, the numerical and experimental results are compared. To make a complete comparison, results obtained with a conventional CP FE model are also included. This model and its parameters are adopted from the original study by Du et al.\ \cite{du2018}. 

In the DSP model, a bandwidth of $\bandwidth=0.8$ $\mu$m is adopted, which is determined through a refinement study similar to that proposed in Wijnen et al.\ \cite{wijnen2021}. The gauge section of the sample is discretized by 3456 quadratic hexahedral elements with 20 nodes and a reduced Gaussian quadrature scheme with 8 quadrature points. A constant displacement is applied to both ends of the sample. Furthermore, the lateral displacement ($y$/$z$-directions) is constrained at both ends, since in the experiment the deformation of the specimen ends is constrained due to the specimen shape \cite{du2018}. This means that all three degrees of freedom are fully prescribed on the top and bottom faces. In earlier studies, these lateral boundary conditions were found to influence the slip activity \cite{du2018, wijnen2021}. 

\begin{figure}[tb!]
	\centering
	\begin{subfigure}{\linewidth}
		\includegraphics[width=\linewidth]{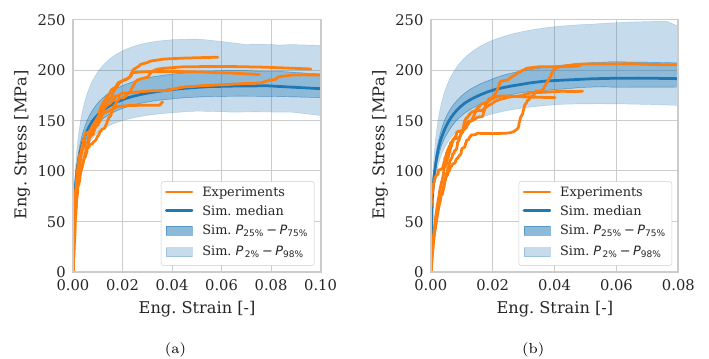}
		\phantomcaption \label{fig:stress_strain_a}
		\phantomcaption \label{fig:stress_strain_b}
	\end{subfigure}
	\caption{Experimental and numerical stress and strain curves for (a) grain 1 and (b) grain 2. The numerical stress-strain curves are represented by their median, 25\%-75\% percentile, and 2\%-98\% percentile.}
	\label{fig:stress_strain}
\end{figure}

The initial slip resistances in the DSP model are determined by \Cref{eq:initialcrss}, where $\latticefriction$, $\shearmodulus$, and $\burgersvector$ are known material parameters which are adopted from literature, while the initial dislocation density, $\dislocationdensity$, of the samples has been measured to be $7\cdot10^{11}$ m$^{-2}$ \cite{du2018}. The nucleation stress, $\nucleationstress$ is sampled from the distribution $\nucleationpdf$. All the identified simulation parameters are given in \Cref{tab:parameters}. The hardening parameters in \Cref{eq:hardening} are identified from the stress-strain curves of the ferrite microtensile tests, such that the median curves of the experiments and simulations adequately match. In \Cref{fig:stress_strain} the experimental stress-strain curves are plotted for Grain 1 (\Cref{fig:stress_strain_a}) and Grain 2 (\Cref{fig:stress_strain_b}) as solid orange lines. The median curve of 100 stress-strain curves obtained with the DSP model is shown as the solid blue line in \Cref{fig:stress_strain}, while the 25\%-75\% and 2\%-98\% percentiles are depicted as, respectively, the dark-blue and bright-blue shaded areas. In the hardening regime, all experimental curves are distributed around the numerical median, and mostly in between the 2\%-98\% percentiles. Only one curve of Grain 2 has a part that deviates significantly from the numerical stress-strain curves due to a large plateau. The degree of scatter between numerical stress-strain curves is mainly determined by $\sourcechance$ (\Cref{eq:sourcechance}), which depends only on physical parameters such as the dislocation density. Adopting a dislocation density one order higher or lower than the measured value would give, respectively, too little or too much scatter.

\begin{figure}[!tb]
	\centering 
	\begin{subfigure}{.95\linewidth}
		\includegraphics[width=\linewidth]{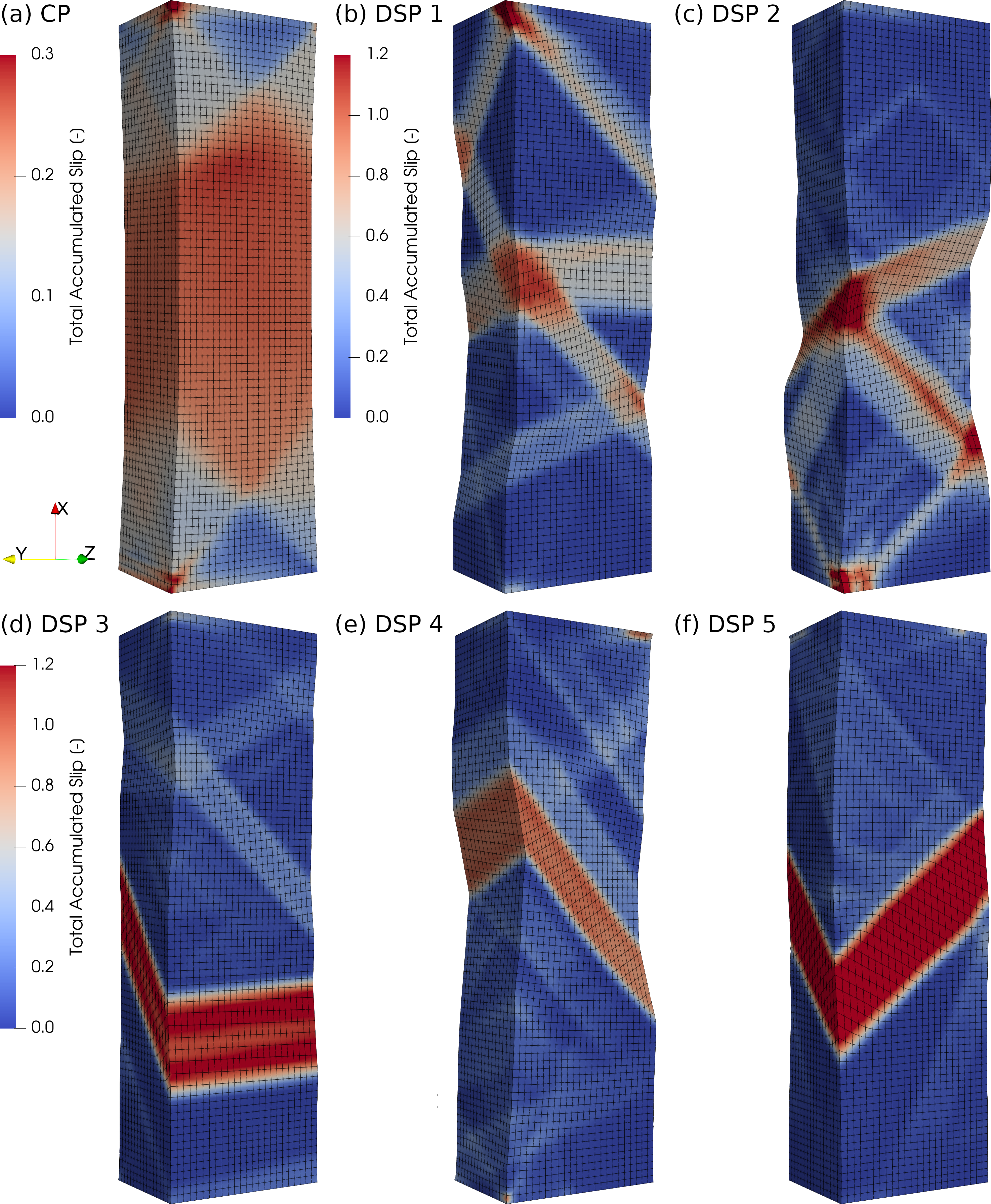}
		\phantomcaption \label{fig:eqstrain_cp} 
		\phantomcaption \label{fig:eqstrain_dsp1}
		\phantomcaption \label{fig:eqstrain_dsp2} 
		\phantomcaption \label{fig:eqstrain_dsp3}
		\phantomcaption \label{fig:eqstrain_dsp4} 
		\phantomcaption \label{fig:eqstrain_dsp5}
	\end{subfigure}
	\caption{Deformed specimens and total accumulated slip contours for (a) the CP model and (b-f) five random realizations of the DSP model for Grain 1.}
\end{figure}

The deformation and total accumulated slip contours obtained with the conventional CP model for Grain 1 are shown in \Cref{fig:eqstrain_cp}. The deformation is smeared out over most of the gauge section, contrary to what is experimentally observed (e.g.\ \Cref{fig:sliptraces}). Close to the boundaries at the top and bottom, passive regions are observed. These regions are the result of the applied boundary conditions, which prohibit deformation of top and bottom surfaces.

Due to the non-deterministic nature of the DSP model, directly comparing a single simulation with a single experiment is not meaningful. Therefore, 100 simulations were done for each grain. This enables a statistical analysis of the simulation data. The deformation of five random realizations simulated with the DSP model is depicted in \Cref{fig:eqstrain_dsp1,fig:eqstrain_dsp2,fig:eqstrain_dsp3,fig:eqstrain_dsp4,fig:eqstrain_dsp5}. Multiple localizations of different slip systems can be noticed, making the deformation of the samples qualitatively similar to the experimental sample of \Cref{fig:sliptraces}, in the sense that localized discrete slip events are captured. Realizations 3, 4, and 5 (\Cref{fig:eqstrain_dsp3,fig:eqstrain_dsp4,fig:eqstrain_dsp5}, respectively) reveal a single dominant localization band. However, these bands are the result of slip on three different slip systems, which are the $(\bar2\bar1\bar1)[\bar111]$, $(\bar110)[\bar1\bar11]$ and $(110)[\bar111]$ systems for realizations 3, 4, and 5, respectively. In contrast, realizations 1 and 2 (\Cref{fig:eqstrain_dsp1,fig:eqstrain_dsp2}, respectively) reveal multiple localization bands of similar magnitude.

To study the influence of non-Schmid effects on the slip activity in the microtensile tests of ferrite, simulations with three different sets of non-Schmid parameters have been performed. In the first parameter set, non-Schmid effects are not taken into account ($\nonschmidcoeff_1=\nonschmidcoeff_2=\nonschmidcoeff_3=0$). In this way, slip is only dependent on the resolved shear stress on the slip plane, in the slip direction, similar to conventional CP models. The second parameter set, which does contain (non-zero) non-Schmid parameters, is adopted from Patra et al.\ \cite{patra2014}, who studied the temperature dependence of the non-Schmid parameters for ferrite, based on experiments performed by Spitzeg \& Keh \cite{spitzig1970}. Their parameters obtained at room temperature are adopted here: $\nonschmidcoeff_1=0.0363$, $\nonschmidcoeff_2=0.1601$, $\nonschmidcoeff_3=0.3243$. Note that these parameters are nearly equal to those found by Mapar et al.\ \cite{mapar2017} in experiments on ferrite samples extracted from dual-phase steel. Furthermore, parameter $\nonschmidcoeff_1$, responsible for the T/AT effect, is approximately zero. This is in agreement with the ab-initio calculations of Dezerald et al.\ \cite{dezerald2016}, who found that the T/AT effect in ferrite is negligible. The third parameter set is adopted from Chen et al.\ \cite{chen2013} ($\nonschmidcoeff_1=0.4577$, $\nonschmidcoeff_2=0.1454$, $\nonschmidcoeff_3=0.5645$). These parameters were identified from atomistic simulations with a magnetic bond-order potential. Contrary to the second set of parameters, the T/AT effect is prominent here.

\subsection{Quantitative slip analysis in different slip directions}
\label{section:results_direction}

At a more quantitative level, first, the slip activity in the different slip directions is considered, without distinguishing the individual slip systems, because the slip activity in the slip directions can be determined with relatively high certainty, as explained in \Cref{section:exp_analysis}. The normalized slip magnitudes in the slip direction are determined per grain and the results of the experiments and the simulations are compared. 

\begin{figure}[!tb]
	\centering
	\begin{subfigure}{\linewidth}
		\includegraphics[width=\linewidth]{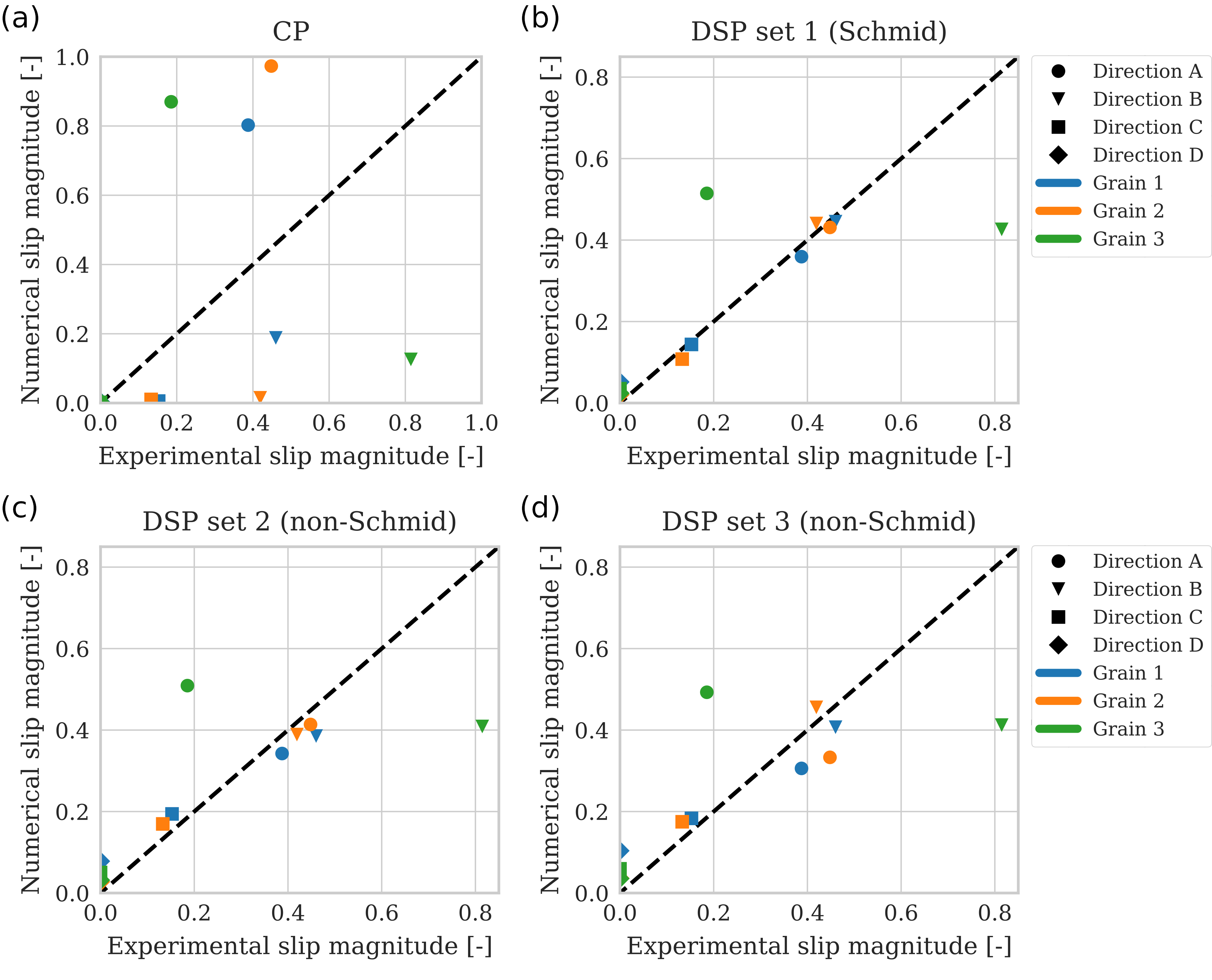}
		\phantomcaption \label{fig:activity_slipdir_conv}
		\phantomcaption \label{fig:activity_slipdir_schmid}
		\phantomcaption \label{fig:activity_slipdir_patra}
		\phantomcaption \label{fig:activity_slipdir_chen}
	\end{subfigure}
	\caption{Normalized slip magnitudes in the four BCC slip directions in the simulations versus the normalized slip magnitudes in all experiments for (a) the CP model and (b) the DSP model with parameter set 1 (Schmid based glide), (c) parameter set 2 and (d) parameter set 3. Slip directions A and D are the slip directions that are, respectively, closest to and furthest away from the 45$\degree$ orientation relative to the loading direction. The dashed line indicates a perfect match between experimental and numerical results.} 
	\label{fig:activity_slipdir}
\end{figure}

First, the results obtained with the conventional CP model are considered, which are shown in \Cref{fig:activity_slipdir_conv}. Here, the calculated normalized slip magnitudes calculated in the simulations are plotted against the normalized slip magnitudes in the experiments. A perfect match between the experimental and numerical slip magnitudes implies that all data points would lie on the dashed line with slope 1. Clearly, the normalized slip magnitudes obtained with the conventional CP model do not agree well with the experimental observations. The mean squared error (MSE) of the data points with respect to the slope 1 line is 0.139. In the CP simulations slip direction \textit{A} is dominant in all grains (i.e.\ slip fraction close to unity), while the experimental slip magnitudes for this slip direction are much lower. This already shows that the conventional CP model is not capable of predicting the global slip system activity in these small scale specimens. The conventional CP model predicts that almost all slip takes place on the primary slip system. Some activity on secondary slip systems is observed due to boundary constraints, which was also demonstrated by Du et al.\ \cite{du2018}. However, the amount of slip on the secondary slip systems observed in the experiments is significantly higher.

The results for the DSP model using parameter set 1 are compared with the experimental results in \Cref{fig:activity_slipdir_schmid}. The experimental and numerical normalized slip magnitudes of Grain 1 (blue data points) and Grain 2 (orange data points) match almost perfectly. However, a large discrepancy between the slip magnitudes of Grain 3 (green data points) is observed. The slip magnitude in direction \textit{A} is significantly higher in the simulations compared to the experiments, while the slip magnitude in direction \textit{B} is significantly lower in the simulations. Nevertheless, the DSP results match the experiments substantially better than for the conventional CP model, revealing an MSE of 0.022. With the discrete slip plane model, more slip takes place on the secondary slip systems with slip direction \textit{B} and even on slip systems with slip direction \textit{C}, because there are not always enough weak dislocation sources present on the primary slip system. This shows the importance of taking the stochastics of dislocation sources into account.

The results of parameter sets 2 and 3, which both consider non-Schmid effects, are shown in, respectively, \Cref{fig:activity_slipdir_patra,fig:activity_slipdir_chen}. Minor differences with respect to the results obtained with parameter set 1, with only Schmid glide, are observed. The obtained MSEs of parameter sets 2 and 3 are 0.024 and 0.025, respectively. The match for slip directions \textit{A} and \textit{B} for Grains 1 and 2 are slightly worse than with parameter set 1, especially for parameter set 3. The large discrepancy between the experimental and numerical normalized slip magnitudes for Grain 3 is present for all three parameter sets. A possible explanation is the small experimental batch size, which consists of only four samples.

\subsection{Quantitative analysis of slip system activity}

We proceed to study the activity of individual slip systems. Recall that these activities, as determined from the experiments, have more uncertainty than those aggregated by slip direction (\Cref{section:results_direction}). It is nevertheless instructive to compare the numerically predicted and experimentally determined amount of slip in detail. We do this here for Grain 2. This grain is considered because it is oriented in single slip, meaning that it has one distinct primary slip system. However, significant slip on secondary slip systems is observed in the experiments, thus revealing the largest discrepancy with the CP simulations. Similar trends can be observed in Grain 1 and Grain 3, the data of which are available in the supplementary material and are not discussed in detail here.

In the simulations, 12 $\slipfamilyA$ and 12 $\slipfamilyB$ slip systems are taken into account. In the experiments, a trace of a $\slipfamilyC$ slip system is occasionally observed, e.g.\ the $(31\bar2)[1\bar11]$ trace in \Cref{fig:sliptraces}. To be able to compare the results, the amount of slip measured on $\slipfamilyC$ slip systems is projected onto the adjacent $\slipfamilyA$ and $\slipfamilyB$ slip systems such that the total slip stays constant. For example, the unit normal of the $(31\bar2)[1\bar11]$ slip system projected onto the unit normals of the $(\bar1\bar10)[1\bar11]$ and $(21\bar1)[1\bar11]$ slip systems yields, respectively, 0.9449 and 0.9820. Hence, fractions of 0.9449/(0.9449+0.9820)=0.4904 and 0.9820/(0.9449+0.9820)=0.5096 of the amount of slip measured on the $(31\bar2)[1\bar11]$ slip system are assigned to the $(\bar1\bar10)[1\bar11]$ and $(21\bar1)[1\bar11]$ slip systems, respectively.

\begin{figure}[tb!]
	\centering
	\begin{subfigure}{\linewidth}
		\centering
		\includegraphics[width=\linewidth]{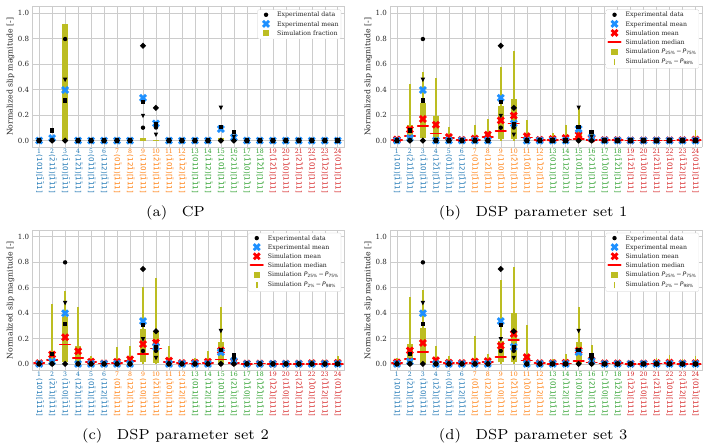}
		\phantomcaption \label{fig:activity_cp}
		\phantomcaption \label{fig:activity_dsp}
		\phantomcaption \label{fig:activity_patra}
		\phantomcaption \label{fig:activity_chen}
	\end{subfigure}
	\caption{Normalized slip magnitudes per slip system for Grain 2 for (a) the CP model and the DSP model with (b)  parameter set 1 (Schmid based glide) (c) parameter set 2 and (d) parameter set 3.. All experiments are plotted as individual data points, together with their mean and median. The DSP simulations are represented by their mean, median, 25\%-75\% percentile and 2\%-98\% percentile.}
	\label{fig:activity_cp_dsp}
\end{figure}

The relative slip system activity for individual slip systems as predicted by the CP model is shown by yellow-green bars in \Cref{fig:activity_cp}. The normalized slip magnitudes in the experiment are plotted as separate data points (black symbols) for each of the five samples, along with the mean of the data points (blue cross). The slip systems are numbered based on their slip direction and orientation. Note that the slip system numbering differs from \Cref{tab:slipfamilyA,tab:slipfamilyB}, where the $\planefamilyA$ and $\planefamilyB$ slip families are grouped. Almost all slip in the simulation takes place on slip system 3, which is the primary slip system. However, in the experiments, a significant amount of slip on slip systems 9, 10, and 15 was also observed. Hence, the CP model is not able to predict the activity of slip systems in the experiments. The MSE between the numerical and experimental mean values is 0.016.

In \Cref{fig:activity_dsp} the slip system activity predicted by the DSP model with only Schmid effects, i.e.\ parameter set 1, is shown. Here, a box plot is used to represent the 100 realizations of the slip resistance distribution. The 2\%, 25\%, 75\%, and 98\% percentiles are used to construct the box plot. Furthermore, the mean of all the realizations is denoted with a red cross. A much more diverse activity of slip systems compared to the CP model is observed. The three slip systems with the highest mean and median, i.e.\ slip systems 3, 9, and 10, are also frequently observed to be active in the experiments. Furthermore, occasional activity on slip systems 2 and 15 is found in both the simulations and the experiments. In the simulations, significant slip activity on slip system 4 is also observed. No slip trace of this slip system is visible in the experiments. To summarize, all slip systems that are observed in the experiments are also active in the DSP simulations, however, one additional slip system is active in the simulations. Nevertheless, a much better agreement with experiments is obtained compared to the CP model. The MSE between the numerical and experimental mean values is equal to 0.0048.

The simulation results for parameter sets 2 and 3, i.e.\ with non-Schmid effects, for Grain 2 are summarized in \Cref{fig:activity_patra,fig:activity_chen}, respectively. Parameter set 2 (\Cref{fig:activity_patra}) does not account for the T/AT effect and has only moderate values for the other two non-Schmid parameters. No significant differences are observed between the slip activity obtained with parameter sets 1 and 2. However, the numerical means of most slip systems are shifted slightly towards the experimental mean, leading to an MSE of 0.0036. This is a small improvement compared to parameter set 1. A more significant difference in slip activity is observed for parameter set 3 (\Cref{fig:activity_chen}). Here, the activity of slip system 4 is almost fully suppressed. This is a result of the A/AT effect, which is prominent in parameter set 3 since slip system 4 is a $\planefamilyB$ slip system that is loaded in its AT direction. All slip systems which are active in simulations for parameter set 3 are active in the experiments and vice versa. However, the MSE of the means is 0.0048, which is equal to the value for parameter set 1 and higher than the value for parameter set 2.

The observations made for Grains 1 and 3 are similar to those for Grain 2. The MSE between the CP model and the experiments is 0.015 for Grain 1 and 0.024 for Grain 3. Note that there is no mean value for the CP simulation since it is a single deterministic result. The MSEs obtained with the mean values of the DSP model without non-Schmid effects (parameter set 1) of both Grains are significantly lower, namely, 0.0042 and 0.0064. No substantial changes in MSEs are observed when non-Schmid effects are taken into account. For parameter set 2, the MSE is 0.0035 for Grain 1 and 0.0060 for Grain 3, which is slightly better compared to parameter set 1. For parameter set 3 the MSE of Grain 1 is 0.0052, which is worse than for the other parameter sets, whereas the MSE of Grain 3 is 0.044, which is slightly better than for the other two parameter sets. The MSEs of all cases discussed above are summarized in \Cref{tab:mse}.

\begin{table}[]
	\centering
	\caption{Mean square errors between the means of the normalized slip magnitudes in the experiments and the simulations. The numbering of the DSP models denotes the parameter set used in the simulations, i.e.\ non-Schmid effects are taken into account in DSP models 2 and 3, but not in DSP model 1.}
	\label{tab:mse}
	\begin{tabular}{@{}ccccc@{}}
		\toprule
		Grain & CP model & DSP model 1 & DSP model 2 & DSP model 3 \\ \midrule
		1     & 0.015    & 0.0042      & 0.0035      & 0.0052      \\
		2     & 0.016    & 0.0048      & 0.0036      & 0.0047      \\
		3     & 0.025    & 0.0064      & 0.0060      & 0.0044      \\ \bottomrule
	\end{tabular}
\end{table}

\subsection{Discussion}

The extensive comparison between experiments and simulations shows that the stochastics of dislocation sources has a major effect on the slip system activity in ferrite at the microscale. The DSP model shows a diversity of active slip systems that is in adequate agreement with the experimental observations, as opposed to the conventional CP model, which mainly shows slip activity on the primary slip system.

Non-Schmid effects seem to only play a minor role in the plastic behavior of ferrite at micrometer scales. Parameter set 2 is considered the most realistic since these experimentally obtained values are reported by two independent studies \cite{patra2014,mapar2017} and comply with the negligible T/AT effect as observed in ab-initio calculations \cite{dezerald2016}. However, no qualitative differences in slip activity are observed with this parameter set compared to the simulations without non-Schmid effects, i.e.\ the same slip systems are active in basically the same proportions. Only a minor improvement in the quantitative slip system activity is obtained by taking this set of non-Schmid parameters into account. When the T/AT effect is fully taken into account, in parameter set 3, the activity of the slip system that is not observed in the experiments, but which is occasionally active in the simulations with the other two parameter sets, is suppressed. In this case, a complete qualitative match of the active slip systems between experiments and simulations is obtained. However, the quantitative match for this parameter set reveals a slightly larger difference than that of parameter set 2.

The above findings suggest that the distribution of dislocation sources, leading to heterogeneous plastic flow, plays a dominant role in the slip system activity of ferrite ingrains. This implies that including the stochastics of dislocation sources in simulations is not only important for capturing the non-smooth strain field but also for accurately predicting the activation of slip systems. Non-Schmid effects, reveal only a minor effect. However, it should be noted that the presented analyses only focus on the slip activity and not on the tension-compression asymmetry of the mechanical response which is often also attributed to non-Schmid effects. Moreover, in specimens of increasing size, the stochastic effects may average out more quickly than the non-Schmid effects, especially in the case of a strong texture. Still, even for larger specimens, stochastic effects deserve attention when the focus lies on predicting localized events, such as the formation of plastic localization bands, the initiation of damage, void nucleation, coalescence, and growth, or even fracture initiation and propagation, for which the DSP model simulations are expected to be valuable.


\section{Conclusions} 
\label{section:conclusions}

In this paper, the slip activity in nanotensile tests of interstitial-free ferrite is studied with the discrete slip plane (DSP) model. To take non-Schmid effects into account, a commonly adopted framework for $\slipfamilyA$ slip systems is extended to $\slipfamilyB$ slip systems. It is shown that the twinning/anti-twinning (T/AT) directions of the $\slipfamilyB$ slip systems are captured by this extension.

An extensive analysis of the active slip systems in the experiments is performed. The results compared the slip activities between the CP model and the DSP model, with and without non-Schmid effects. The main conclusions that are drawn for the behavior of ferrite at the microscale are:
\begin{itemize}
	\item The slip system activity in the experiments critically depends on the heterogeneity of plastic deformation due to the presence of dislocation sources and obstacles.
	\item Non-Schmid effects do not significantly contribute to the slip system activity of ferrite at the microscale.
\end{itemize}
This has the following implications for modeling:
\begin{itemize}
	\item Continuum crystal plasticity models with uniform properties are not suitable for analyzing small-scale mechanical tests since they fail to predict both the discrete character of the displacement field and the diversity of activated slip systems.
	\item Modeling frameworks that do account for the stochastics of the dislocation network, such as the DSP model, should be used to analyze small-scale experiments. Not only the average mechanical behavior of specimens is affected, but also the outliers which might be more critical for localized events, such as the formation of plastic localization bands, damage initiation, void nucleation, and fracture.
\end{itemize}


\section*{Author Contributions (CRediT)}

\noindent
\textbf{Job Wijnen:} Conceptualization, Methodology, Investigation, Formal Analysis, Software, Writing - Original Draft, Visualization \\
\textbf{Johan Hoefnagels:} Conceptualization, Methodology, Resources, Writing - Review \& Editing, Supervision, Funding Acquisition \\
\textbf{Marc Geers:} Methodology, Resources, Writing - Review \& Editing, Supervision, Funding Acquisition \\
\textbf{Ron Peerlings:} Conceptualization, Methodology, Writing - Review \& Editing, Supervision, Funding Acquisition

\section*{Declaration of Competing Interest}

The authors declare that they have no known competing financial interests or personal relationships that could have appeared to influence the work reported in this paper.

\section*{Acknowledgments}

This research was carried out under project number S17012a in the framework of the Partnership Program of the Materials innovation institute M2i (\href{www.m2i.nl}{www.m2i.nl}) and the Netherlands Organization for Scientific Research NWO (\href{www.nwo.nl}{www.nwo.nl}; project number 16348).

Chaowei Du is gratefully acknowledged for providing the experimental data used in this study.

\section*{Data availability}

Details of the experimental analysis are available in the supplementary material to this paper. The raw and/or processed data required to reproduce these findings will be made available upon request.


\bibliography{references.bib}


\clearpage


\appendix

\section{Slip systems and simulation parameters}
\label{appendix:slipsystems}

\begin{table}[h!]
	\centering
	\caption{Orientations of all three grains in Proper Euler angles ($Z_1X_2Z_3$ convention).}
	\label{tab:eulerangles}
	\begin{tabular}{@{}llll@{}}
		\toprule
		Grain & $\phi_1$ [$\degree$] & $\phi_2$ [$\degree$] & $\phi_3$ [$\degree$] \\ \midrule
		1     & 333.6                    & 37.9                    & 25.7                    \\
		2     & 320.1                    & 41.2                    & 42.9                    \\
		3     & 258.7                    & 142.6                   & 258.3                   \\ \bottomrule
	\end{tabular}
\end{table}

\begin{table}[h!]
	\centering
	\caption{Slip plane normals, slip directions and auxiliary projection plane normals for $\slipfamilyA$ slip systems.}
	\label{tab:slipfamilyA}
	\begin{tabular}{@{}lllllllllll@{}}
		\toprule
		$\alpha$ & $\slipplanenormal$ & $\slipdirection$ & $\projectionplanenormal^+$ & $\projectionplanenormal^-$ & $\quad$ & $\alpha$ & $\slipplanenormal$ & $\slipdirection$ & $\projectionplanenormal^+$ & $\projectionplanenormal^-$ \\ \midrule
		1        & $(01\bar1)$        & $\left<111\right>$          & $(\bar110)$                & $(10\bar1)$                &                       & 7        & $(0\bar1\bar1)$    & $\left<\bar1\bar11\right>$  & $(1\bar10)$                & $(\bar10\bar1)$            \\
		2        & $(\bar101)$        & $\left<111\right>$          & $(0\bar11)$                & $(\bar110)$                &                       & 8        & $(101)$            & $\left<\bar1\bar11\right>$  & $(011)$                    & $(1\bar10)$                \\
		3        & $(1\bar10)$        & $\left<111\right>$          & $(10\bar1)$                & $(0\bar11)$                &                       & 9        & $(\bar110)$        & $\left<\bar1\bar11\right>$  & $(\bar10\bar1)$            & $(011)$                    \\
		4        & $(\bar10\bar1)$    & $\left<\bar111\right>$      & $(\bar1\bar10)$            & $(01\bar1)$                &                       & 10       & $(10\bar1)$        & $\left<1\bar11\right>$      & $(110)$                    & $(0\bar1\bar1)$            \\
		5        & $(0\bar11)$        & $\left<\bar111\right>$      & $(101)$                    & $(\bar1\bar10)$            &                       & 11       & $(011)$            & $\left<1\bar11\right>$      & $(\bar101)$                & $(110)$                    \\
		6        & $(110)$            & $\left<\bar111\right>$      & $(01\bar1)$                & $(101)$                    &                       & 12       & $(\bar1\bar10)$    & $\left<1\bar11\right>$      & $(0\bar1\bar1)$            & $(\bar101)$           \\   \bottomrule
	\end{tabular}
\end{table}

\begin{table}[htbp]
	\centering
	\caption{Slip plane normals and slip directions of $\slipfamilyB$ slip systems. Furthermore, the pairs of $\slipfamilyA$ slip systems that together constitute the apparent $\slipfamilyB$ slip system are given.} 
	\label{tab:slipfamilyB}
	\begin{tabular}{@{}llllrcl@{}}
		\toprule
		$\alpha$ & $\slipplanenormal$  & $\slipdirection$           & $\quad$ & $\alpha_1$ & $-$ & $\alpha_2$ \\ \midrule
		13       & $(\bar12\bar1)$     & $\left<111\right>$         &          & 1          & $-$ & 3          \\
		14       & $(\bar1\bar12)$     & $\left<111\right>$         &          & 2          & $-$ & 1          \\
		15       & $(2\bar1\bar1)$     & $\left<111\right>$         &          & 3          & $-$ & 2          \\
		16       & $(\bar2\bar1\bar1)$ & $\left<\bar111\right>$     &          & 4          & $-$ & 6          \\
		17       & $(1\bar12)$         & $\left<\bar111\right>$     &          & 5          & $-$ & 4          \\
		18       & $(12\bar1)$         & $\left<\bar111\right>$     &          & 6          & $-$ & 5          \\
		19       & $(1\bar2\bar1)$     & $\left<\bar1\bar11\right>$ &          & 7          & $-$ & 9          \\
		20       & $(112)$             & $\left<\bar1\bar11\right>$ &          & 8          & $-$ & 7          \\
		21       & $(\bar21\bar1)$     & $\left<\bar1\bar11\right>$ &          & 9          & $-$ & 8          \\
		22       & $(21\bar1)$         & $\left<1\bar11\right>$     &          & 10         & $-$ & 12         \\
		23       & $(\bar112)$         & $\left<1\bar11\right>$     &          & 11         & $-$ & 10         \\
		24       & $(\bar1\bar2\bar1)$ & $\left<1\bar11\right>$     &          & 12         & $-$ & 11         \\ \bottomrule
	\end{tabular}
\end{table}

\begin{table}[htbp]
	\centering
	\caption{Parameters adopted in the simulations}
	\label{tab:parameters}
	\begin{tabular}{@{}lll@{}}
		\toprule
		Model parameter              & Symbol             & Value                    \\ \midrule
		Lattice constant             & $a$                & 2.856 \r{A}      \\
		Lattice friction             & $\latticefriction$ & 10 MPa                   \\
		Saturation CRSS              & $\saturationcrss$  & 4$\crss$                 \\
		Hardening rate               & $\hardeningrate$   & 30 GPa/$\mu$m            \\
		Hardening exponent           & $\hardeningexp$    & 5                        \\
		Latent hardening coefficient & $q_n$              & 1.4                      \\
		Initial dislocation density  & $\rho_0$           & $7\cdot10^{11}$ m$^{-2}$ \\
		Rate sensitivity parameter   & $m$                & 0.05                     \\
		Reference velocity           & $\dot{v}_0$        & 9$\cdot10^{-2}$ $\mu$m/s \\
		Elastic constants            & $C_{11}$           & 233.5 MPa                \\
									 & $C_{12}$           & 135.5 MPa                \\
									 & $C_{44}$           & 118.0 MPa                \\ 
		Non-Schmid parameter set 1  & $\nonschmidcoeff_1$ & 0 \\
		                            & $\nonschmidcoeff_2$ & 0 \\
		                            & $\nonschmidcoeff_3$ & 0 \\
	    Non-Schmid parameter set 2  & $\nonschmidcoeff_1$ & 0.0363 \\
								    & $\nonschmidcoeff_2$ & 0.1601 \\
								    & $\nonschmidcoeff_3$ & 0.3243 \\
	    Non-Schmid parameter set 3  & $\nonschmidcoeff_1$ & 0.4577 \\
								    & $\nonschmidcoeff_2$ & 0.1454 \\
								    & $\nonschmidcoeff_3$ & 0.5645 \\
		\bottomrule
	\end{tabular}
\end{table}


\end{document}